\documentclass[a4paper,11pt]{article}
\usepackage{pos}
\usepackage{natbib}
\usepackage{lineno}

\def\pt{P_{\rm{T}}}

\def\gevc2{\rm{GeV}/\textit{c}^2}

\def\zu{\hat{\textbf{z}}}

\def\pythia{\texttt{PYTHIA}}
\def\3p0{string+${}^3P_0$}
\def\StringSpinner{\texttt{StringSpinner}}
\def\kpt{\textbf{k}'_{\rm{T}}}
\def\kptkpt{\textbf{k}'^2_{\rm{T}}}

\def\Sqt{\textbf{S}_{q\rm{T}}}
\def\Im{\rm{Im}}
\def\fL{f_{\rm{L}}}
\def\GL{G_{\rm{L}}}
\def\GT{G_{\rm{T}}}
\def\thetaLT{\theta_{\rm{LT}}}
\def\D{\check{D}}
\def\vecsigma{\boldsymbol{\sigma}}

\title{Adding quark spin effects to \texttt{PYTHIA} string fragmentation}

\author*[a]{Albi Kerbizi}
\affiliation[a]{Trieste Section of INFN, University of Trieste, Dept. of Physics, 34127 Trieste, Italy}
\emailAdd{albi.kerbizi@ts.infn.it}

\author[b]{Leif L\"onnblad}
\affiliation[b]{Department of Astronomy and Theoretical Physics, Lund University, Sweden}
\emailAdd{leif.lonnblad@thep.lu.se}

\abstract{Quark spin effects in hadronization have been recently included in the \texttt{PYTHIA 8} event generator for the simulation of the deep inelastic scattering (DIS) process off a polarized nucleon target via the external \StringSpinner{} package. The spin effects can be simulated for the production of pseudoscalar mesons using the string+${}^3P_0$ model of polarized quark fragmentation and parametrizations of the transversity PDFs.

We present here a major development of \StringSpinner{} which includes the production of vector mesons in the polarized \texttt{PYTHIA 8} string fragmentation as well as their polarized decays. The package is validated and used to simulate the transverse-spin asymmetries in semi-inclusive DIS (SIDIS). The simulation results on the Collins and dihadron asymmetries for the final state mesons are compared with the asymmetries measured in SIDIS off transversely polarized protons. Also shown are the predictions for the Collins asymmetries for $\rho$ mesons produced in SIDIS off transversely polarized protons.}

\FullConference{%
  41st International Conference on High Energy physics - ICHEP2022\\
  6-13 July, 2022\\
  Bologna, Italy
}

\begin{document}
\maketitle

\section{Introduction}\label{sec:Intro}
Monte Carlo event generators (MCEGs) are fundamental tools in high energy physics commonly used for the extraction of physics information from data, for the interpretation of data and to make predictions for future experiments. The requirements for the construction of MCEGs come mainly from the LHC physics program while other interesting phenomena explored in recent years are typically not considered. The most remarkable example is the observation of the Collins transverse spin asymmetry (TSA) in semi-inclusive DIS (SIDIS) processes \cite{Hermes:Collins, Compass:Collins} and the corresponding azimuthal asymmetry for back-to-back hadrons produced in $e^+e^-$ annihilation \cite{Belle:2019}. 
The combined analysis of these asymmetries led to the extractions of the transversity PDF, describing the transverse polarization of a quark in a transversely polarized nucleon, and to the Collins fragmentation function (FF), describing the fragmentation of a transversely polarized quark in hadrons (the so-called Collins effect \cite{Collins:1993}). Other examples are the dihadron transverse-spin asymmetry \cite{Compass:Dihadron} and the jet-handedness \cite{Jlab:Beam}. 

To enable the simulation of these effects with a MCEG, we started a systematic introduction of quark spin effects in the string fragmentation routine of the \texttt{PYTHIA 8} event generator \cite{Sjostrand:2008} with the \StringSpinner{} package \cite{Kerbizi:StringSpinner}. The spin effects can be simulated for the DIS process off a transversely polarized nucleon with pseudoscalar (PS) meson production using the recent string+${}^3P_0$ model of polarized quark fragmentation \cite{Kerbizi:2021} and parametrizations of the transversity PDFs for the calculation of the transverse polarization of the struck quark. The model is capable of reproducing the Collins effect as well as the dihadron and jet handedness effects. Parton showers are switched off, as their activation would require further developments of the \3p0{} model to handle the fragmentation of the complex string topologies produced by the showering process.

In this article we present a major recent development of \texttt{StringSpinner}, namely the production and decay of vector mesons (VMs) in the final states of DIS events. To achieve this goal we used the most recent version of the \3p0{} model as well as its standalone MC implementation \cite{Kerbizi:2021}. This development allows for a more complete simulation of transverse spin effects in DIS events and it is described in Sec. \ref{sec:Incl}. The new version of \StringSpinner{} is used to compare the simulated TSAs with the data from SIDIS measurements and the results are shown in Sec. \ref{sec:Comp}. In Sec. \ref{sec:Coll} we show the simulation results for the Collins asymmetries for VM production in SIDIS, another interesting observable arising in the polarized fragmentation process which is still poorly explored theoretically and experimentally. Finally in Sec. \ref{Sec:Conclusions} we draw our conclusions.

\section{Inclusion of spin effects in \pythia{} hadronization}\label{sec:Incl}
The introduction of quark spin effects in the string fragmentation process of \pythia{} is achieved using the \texttt{UserHooks} class of the generator. This class allows an external user to step in during the execution of the normal string fragmentation process and to veto each produced hadron according to some externally imposed logic. The logic implemented in \StringSpinner{} is inspired to the \3p0{} model presented in Ref. \cite{Kerbizi:2021} and it is the following.

Each hadron $h$ emitted in string fragmentation is viewed as arising from the splitting $q\rightarrow h + q'$, where $q$ is the fragmenting quark and $q'$ the leftover quark. The hadron is accepted with the probability
\begin{equation}\label{eq:probability}
    p(\kpt,\Sqt) = \frac{1}{2} \times \left[ 1 + a \frac{2\Im(\mu)}{|\mu|^2+\kptkpt} \, \Sqt\cdot \left(\zu\times \kpt\right) \right].
\end{equation}
The $\zu$ axis defines the string axis, stretched in a DIS event between the struck quark and the target remnant. The vectors $\kpt$ and $\Sqt$ are the transverse momentum of $q'$ and the transverse polarization of $q$ with respect to the string axis, respectively. The vector $\Sqt$ is encoded in the spin density matrix $\rho(q)$ of $q$. The quantity $\mu$ is a complex mass parameter introduced in the \3p0{} model to parametrize the relative ${}^3P_0$ wave function of the $q'\bar{q}'$ pair produced at the string breaking. The imaginary (real) part is responsible for transverse (longitudinal) spin effects in the fragmentation process.

The factor $a$ is constant, and it is $a=-1$ for a PS meson and $a=\fL$ for a VM. $\fL$ is a free parameter of the \3p0{} model which gives the fraction of longitudinally polarized VMs w.r.t the string axis. It is defined as $\fL=|\GL|^2/(2|\GT|^2+|\GL|^2)$, where $\GL$ and $\GT$ are the complex constants describing the coupling of quarks to VMs with longitudinal and transverse polarization with respect to the string axis. The combination of the coupling constants $\thetaLT=\arg(\GL/\GT)$ defines the second free parameter of the \3p0{} model needed to describe the spin effects for VM production. A non-zero $\thetaLT$ is responsible for the oblique polarization of VMs \cite{Kerbizi:2021}.

If a VM is accepted after the veto procedure, its spin density matrix is calculated using \cite{Kerbizi:2021}
\begin{eqnarray}
    \rho_{\alpha\alpha'}(h) = \frac{ \rm{Tr} \left[(\mu + \sigma_z\, \vecsigma\!\cdot\!\kpt) \,\Gamma_{h,\alpha} \, \rho(q) \, 
\Gamma^\dag_{h,\alpha'} \, (\mu^* +  \vecsigma\!\cdot\!\kpt \, \sigma_z) \right] } { \rm{Tr} \left[(\mu + \sigma_z\, \vecsigma\!\cdot\!\kpt) \,\Gamma_{h,\beta} \, \rho(q) \, 
\Gamma^\dag_{h,\beta} \, (\mu^* +  \vecsigma\!\cdot\!\kpt \, \sigma_z) \right] },
\end{eqnarray}
where $\Gamma_{h,\alpha} = \left(\GT\sigma_x\sigma_z, \GT\sigma_y\sigma_z,\GL\,1_{2\times 2}\right)$ defines the coupling of quarks to the VM and $\alpha=x,y,z$. The $\sigma$ matrices indicate the Pauli matrices. The density matrix $\rho(h)$ is used to generate the polarized hadronic decays of the VM as explained in Ref. \cite{Kerbizi:2021} using as external routines those developed in the stand alone implementation of the \3p0{} model. The decay products of the VM are book-kept and provided to \pythia{} at at later stage (see below).

The external decays also give in output a decay matrix $\D$, encoding the correlations between the polarization of the fragmenting quark $q$ and the decay products of the VM. This matrix is used to calculate the density matrix of the leftover quark $q'$
\begin{equation}
     \rho(q') = \left[
\D_{\alpha\alpha'}\,(\mu + \sigma_z\, \vecsigma\cdot\kpt) \, \Gamma_{h,\alpha} \, \rho(q) \, 
\Gamma^\dag_{h,\alpha'} \, (\mu^* +  \vecsigma\cdot\kpt \, \sigma_z)\right] / \rm{Tr}\left[ Numerator \right],
\end{equation}
which allows to propagate the spin-correlations along the fragmentation chain \cite{Collins:SpinCorr}, 
until the chain is ended by \pythia{}.

After string fragmentation, \pythia{} starts the decays of the produced resonances. The methods of the \texttt{UserHooks} class are again used by \StringSpinner{} to provide \pythia{} with the decay products of the externally simulated decays, which are finally stored in the event record. Non-hadronic decay channels of VMs as well as all the decays of PS mesons are handled by \pythia{}.

The introduction of quark spin effects in the string fragmentation and decay routines of \pythia{} is complemented by the calculation of the polarization of the struck quark in a DIS event, which is performed as in Ref. \cite{Kerbizi:StringSpinner}.

\section{Comparison with transverse-spin asymmetries in SIDIS}\label{sec:Comp}
The new version of \StringSpinner{} is validated by comparing the relevant simulation results with those obtained from the standalone MC implementation of the \3p0{} model \cite{Kerbizi:2021}, finding the very same results (not shown here). 

\begin{figure}[tbh]
\centering
\begin{minipage}[b]{0.56\textwidth}
\includegraphics[width=1.015\textwidth]{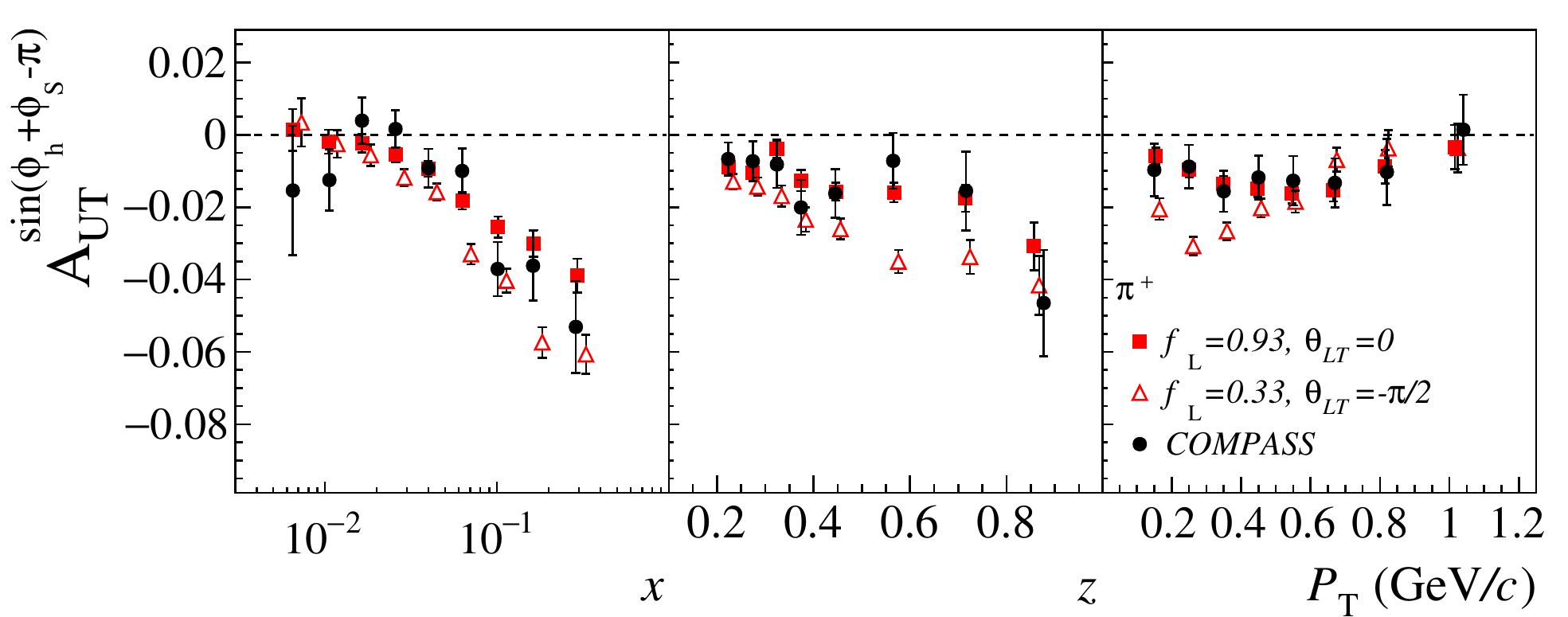}
\end{minipage}
\begin{minipage}[b]{0.56\textwidth}
\includegraphics[width=1.015\textwidth]{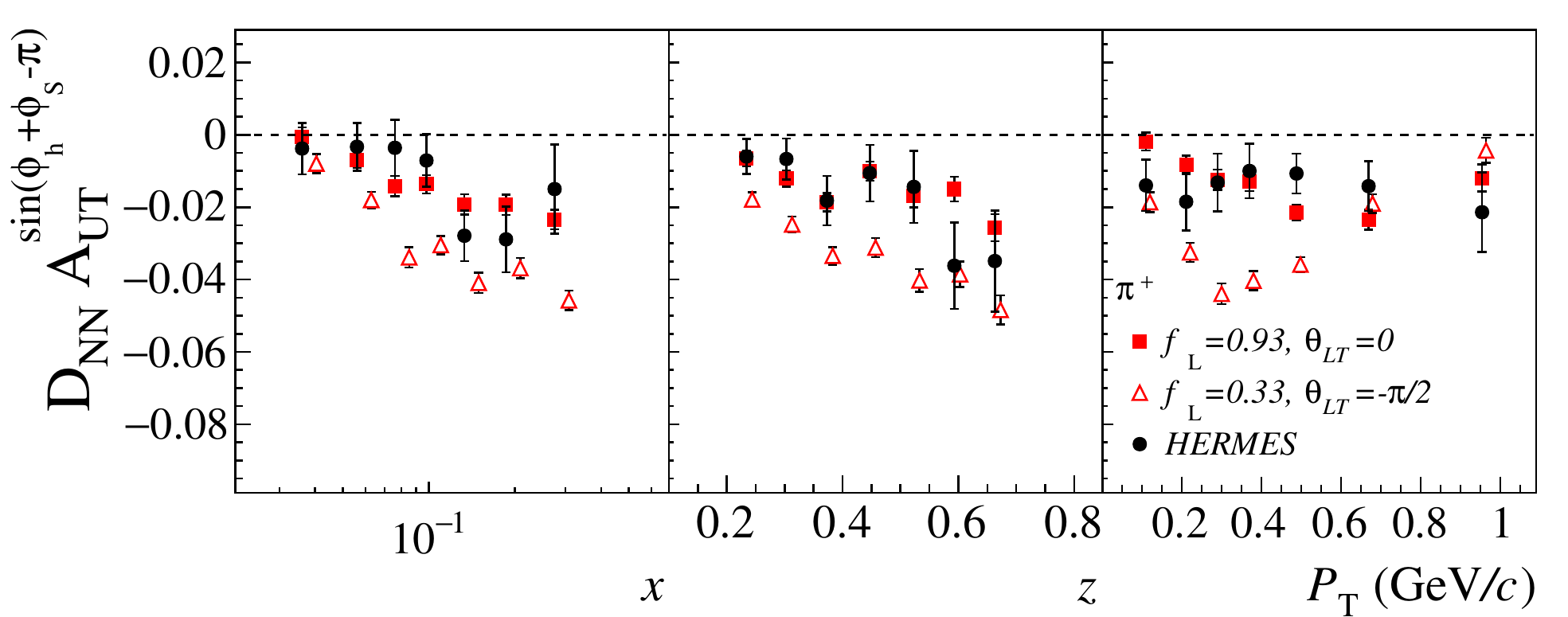}
\end{minipage}
\caption{Top row: Comparison between the simulated Collins asymmetry (squares and triangles) and the asymmetry measured by COMPASS \cite{Compass:Collins} (circles) for $\pi^+$ as function of $x$ (left), $z$ (middle) and $\pt$ (right). Bottom row: comparison between the simulated Collins asymmetry multiplied by the quark depolarization factor $\rm{D_{NN}}$ and the HERMES data \cite{Hermes:Collins} (the same sign convention as in the upper row is used).}
\label{fig:Collins pi+}
\end{figure}

Figure \ref{fig:Collins pi+} shows the comparison between the Collins asymmetry $\rm{A_{UT}^{\sin(\phi_h+\phi_ S-\pi)}}$ for $\pi^+$ as obtained from simulations of the SIDIS process off transversely polarized protons with the COMPASS data \cite{Compass:Collins} (top row) and HERMES data \cite{Hermes:Collins} (bottom row). The asymmetry $\rm{A_{UT}^{\sin(\phi_h+\phi_S-\pi)}}$ is defined as the amplitude of the $\sin(\phi_{\rm h}+\phi_{\rm S}-\pi)$ modulation in the SIDIS cross section, where $\phi_{\rm h}$ and $\phi_{\rm S}$ are the azimuthal angles of the pion and of the target transverse polarization around the exchanged virtual photon momentum, respectively. The same kinematic cuts as in the real data analyses are applied. In each panel the asymmetries are shown as function of the Bjorken-$x$ variable, the fraction $z$ of the available energy carried by the $\pi^+$ and its transverse momentum $\pt$ with respect to the momentum of the exchanged virtual photon. The values of the free parameters used in simulations are the same as in Ref. \cite{Kerbizi:2021}. The simulation results are shown for the two choices $\fL=0.93$ and $\thetaLT=0$ (squares), and $\fL=0.33$ and $\thetaLT=-\pi/2$ (triangles). The former choice favours the production of longitudinally polarized VMs without oblique polarization, and it gives a satisfactory description of both COMPASS and HERMES data. The latter choice, instead, gives equal probabilities for the production of longitudinally and transversely polarized VMs and allows for a non-zero oblique polarization, giving an unsatisfactory description of the data. 
Other combinations of $\fL$ and $\thetaLT$ which lead to a satisfactory description of data are also possible \cite{Kerbizi:2021}. In fact, more precise SIDIS data would be needed to determine the values of these free parameters. Concerning the asymmetries for $\pi^-$, which depend weakly on $\fL$ and $\thetaLT$, a satisfactory description of data is obtained as well.

\begin{figure}[tbh]
\centering
\begin{minipage}[b]{0.6\textwidth}
\centering
\includegraphics[width=1.015\textwidth]{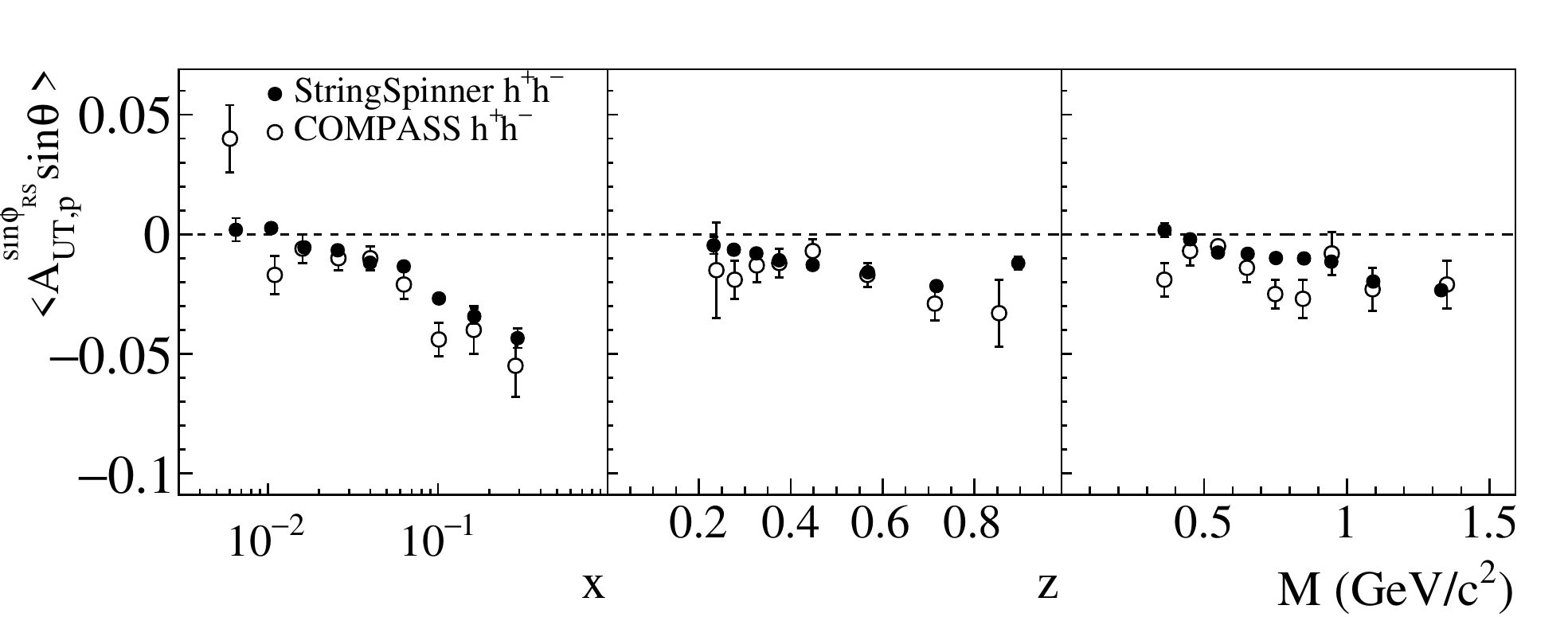}
\end{minipage}
\caption{Comparison between the simulated dihadron asymmetry for $h^+h^-$ pairs (full points) and the asymmetry measured by COMPASS \cite{Compass:Dihadron} as function of $x$ (left), $z$ (middle) and invariant mass $M$ (right). 
}
\label{fig:Dihadron}
\end{figure}

Simulations have also been carried out to calculate the dihadron TSA $\rm{\langle A_{UT,p}^{\sin\phi_{ RS}}\,\sin\theta\rangle}$ for $h^+h^-$ pairs in SIDIS off transversely polarized protons. The asymmetry $\rm{\langle A_{UT,p}^{\sin\phi_{ RS}}\,\sin\theta\rangle}$ is the amplitude of the $\rm{\sin(\phi_R+\phi_S-\pi)}$ modulation in the cross section for the production of a $h^+h^-$ pair, $\phi_{R}$ being the azimuthal angle of the relative momentum of the pair around the direction of the exchanged virtual photon \cite{Compass:Dihadron}. The result obtained with $\fL=0.93$ and $\thetaLT=0$ for the COMPASS kinematics is shown in Fig. \ref{fig:Dihadron} (closed circles) as function of $x$, the fraction $z$ of the available energy carried by the pair and the invariant mass $M$ of the pair. A satisfactory comparison with the COMPASS data (empty points) is found as function of the explored kinematic variables. Other choices of $\fL$ and $\thetaLT$ have a weak effect on the dihadron TSA, meaning that this asymmetry is not suitable for the tuning of such parameters.

\vspace{-2.5em}
\section{Collins asymmetries for $\rho$ mesons in SIDIS}\label{sec:Coll}
\begin{figure}[tbh]
\centering
\begin{minipage}[b]{0.56\textwidth}
\includegraphics[width=1.05\textwidth]{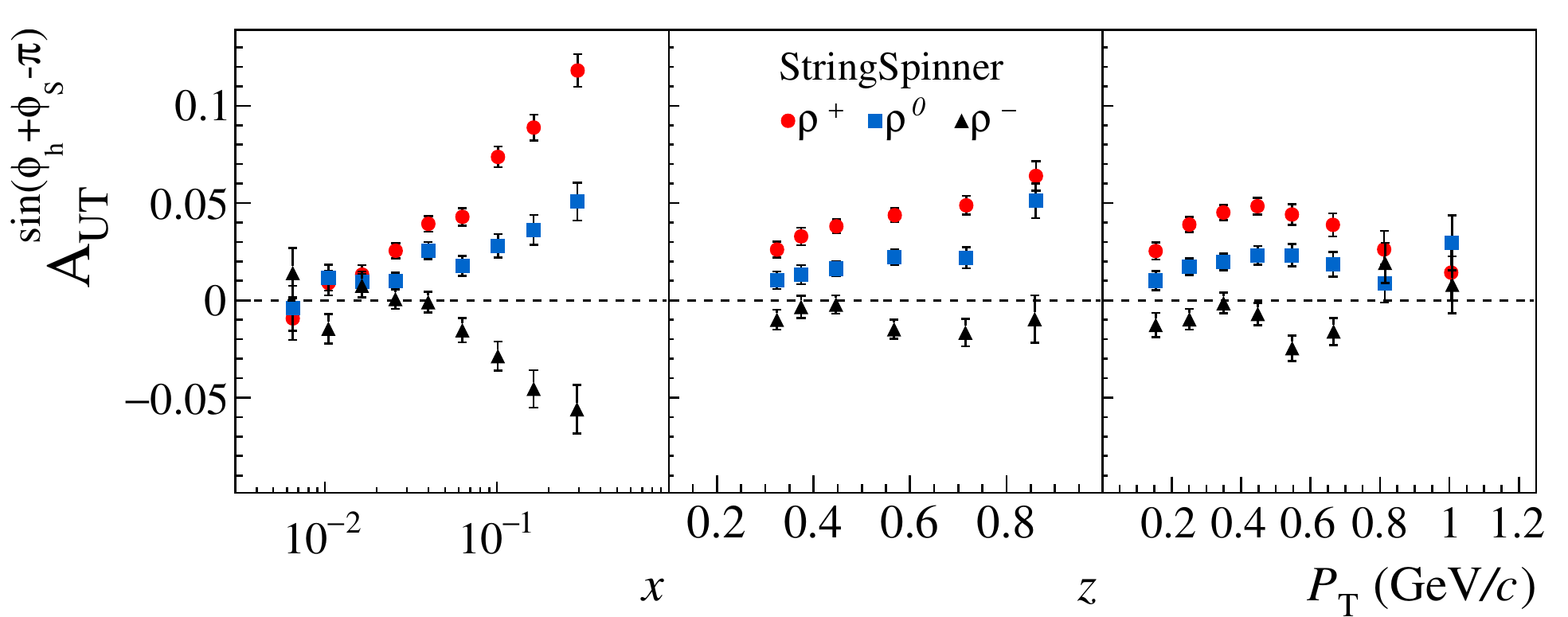}
\end{minipage}
\begin{minipage}[b]{0.56\textwidth}
\includegraphics[width=1.05\textwidth]{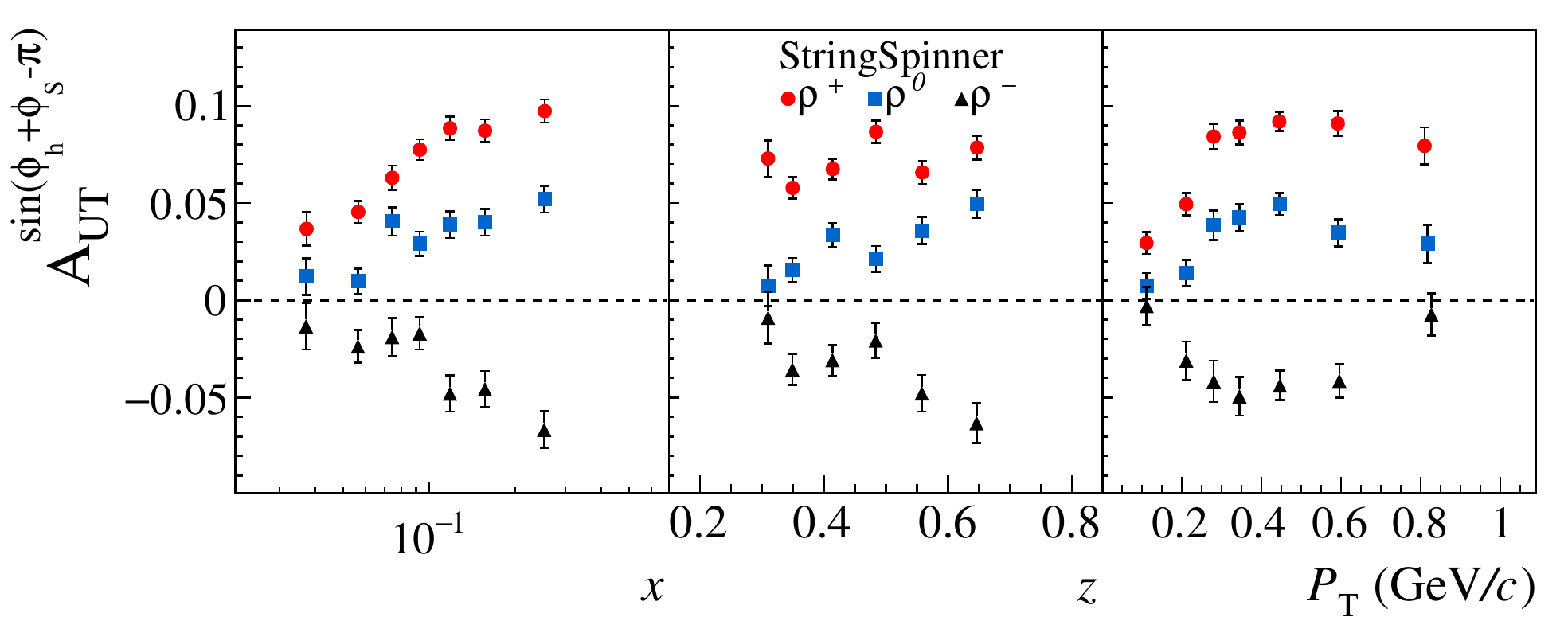}
\end{minipage}
\caption{Top row: the simulated Collins asymmetry for $\rho^+$ (circles), $\rho^0$ (squares) and $\rho^-$ (triangles) in SIDIS off protons in the COMPASS kinematics. Bottom row: the same asymmetries in the HERMES kinematics.}
\label{fig:Collins rho}
\end{figure}

Figure \ref{fig:Collins rho} shows the simulation results for the Collins asymmetries for $\rho$ mesons, as function of $x$, $z$ and $\pt$, in the COMPASS kinematics (top row) and HERMES kinematics (bottom row). The free parameters $\fL=0.93$ and $\thetaLT=0$ are used. Also, the same kinematic requirements as in the COMPASS analysis of the Collins TSA for $\rho^0$ mesons produced in SIDIS off protons are used \cite{Compass:rho0}.

As can be seen, the asymmetry for $\rho^+$ is large and increases up to $10\%$ for large values of $x$, both in the COMPASS and HERMES kinematic configurations. Also, the asymmetry is larger compared to the $\pi^+$ asymmetry shown in Fig. \ref{fig:Collins pi+}. This is partly because of the higher $z$ cut applied to $\rho^+$ mesons and due to the fact that only a small fraction (few percent) of the $\rho^+$ mesons come from decays of heavier resonances, as suggested by \pythia{}. Concerning $\rho^-$ mesons, their asymmetry is smaller in size compared to the $\rho^+$ one, and it is negative. 
Finally, the asymmetry for $\rho^0$ mesons has a size in the between of that for $\rho^+$ and that for $\rho^-$ mesons, as expected from isospin invariance, and has positive value, as indicated by the measurement of the COMPASS experiment \cite{Compass:rho0}.

The Collins asymmetries for $\rho$ mesons have been calculated also for other values of $\fL$ and $\thetaLT$, finding a strong dependence on $\fL$ and essentially no dependence on $\thetaLT$, as expected from Eq. (\ref{eq:probability}). The asymmetry vanishes at small $\fL$, namely when VMs are mainly produced with transverse polarization. The sensitivity on $\fL$ makes the measurements of the Collins asymmetries for $\rho$ mesons suitable for the tuning of this parameter. These measurements are however difficult because of the high combinatorial background in the hadron sample used to identify the $\rho$ mesons, but feasible, as shown in Ref. \cite{Compass:rho0}.

\section{Conclusions} \label{Sec:Conclusions}
We have introduced the quark-spin effects in the hadronization routines of the \texttt{PYTHIA 8} event generator for the simulation of the polarized DIS process with PS meson and VM production. The spin effects are enabled by the \StringSpinner{} package, which uses the \3p0{} model of polarized fragmentation to propagate these effects along the fragmentation chain. The package can be used to simulate transverse spin asymmetries in semi-inclusive DIS. The results of the simulations have been compared with the available data on the Collins and dihadron asymmetries, finding a satisfactory description. A write-up of the new version of \StringSpinner{} is in preparation and is foreseen to be published soon.


\begin{thebibliography}{99}
\bibitem{Hermes:Collins}
A. Airapetian \emph{et al.} (HERMES Collaboration), \href{https://doi.org/10.1016/j.physletb.2010.08.012}{Phys. Lett. B \textbf{693} (2010) 11-16.}

\bibitem{Compass:Collins}
C. Adolph \emph{et al.} (COMPASS Collaboration), \href{https://doi.org/10.1016/j.physletb.2015.03.056}{Phys. Lett. B \textbf{744}, 250 (2015).}

\bibitem{Belle:2019}
H. Li \emph{et al.} (Belle Collaboration), \href{https://doi.org/10.1103/PhysRevD.100.092008}{Phys. Rev. D \textbf{100}, 092008 (2015)}; J. P. Lees \emph{et al.} (BaBar Collaboration) \href{https://doi.org/10.1103/PhysRevD.90.052003}{Phys. Rev. D \textbf{90} (2014) 5, 052003}; M. Ablikim \emph{et al.} (BESIII Collaboration), \href{https://doi.org/10.1103/PhysRevLett.116.042001}{Phys. Rev. Lett. \textbf{116}, 042001}.



\bibitem{Collins:1993}
 J. C. Collins, \href{https://doi.org/10.1016/0550-3213(93)90262-N}{Nucl. Phys. B \textbf{396}, 161 (1993).}

\bibitem{Compass:Dihadron}
C. Adolph \emph{et al.} (COMPASS Collaboration), \href{https://doi.org/10.1016/j.physletb.2014.06.080}{Phys. Lett. B \textbf{736}, 124 (2014).}

\bibitem{Jlab:Beam}
T. B. Hayward \emph{et al.} (CLAS Collaboration), \href{https://doi.org/10.1103/PhysRevLett.126.152501}{Phys. Rev. Lett. \textbf{126}, 152501.}




\bibitem{Sjostrand:2008}
T. Sjostrand, S. Mrenna, and P. Z. Skands, \href{https://doi.org/10.1088/1126-6708/2006/05/026}{J. High Energy
Phys. \textbf{05} (2006) 026}; \href{https://doi.org/10.1016/j.cpc.2008.01.036}{Comput. Phys. Commun. \textbf{178}, 852 (2008).}

\bibitem{Kerbizi:StringSpinner}
A. Kerbizi and L. L\"onnblad, \href{https://doi.org/10.1016/j.cpc.2021.108234}{Comput. Phys. Commun. \textbf{272} (2022) 108234.}

\bibitem{Kerbizi:2021}
A. Kerbizi, X. Artru and A. Martin, \href{https://doi.org/10.1103/PhysRevD.104.114038}{Phys. Rev. D \textbf{104} (2021) 11, 114038.}

\bibitem{Collins:SpinCorr}
 J. C. Collins, \href{https://doi.org/10.1016/0550-3213(88)90654-2}{Nucl. Phys. B \textbf{304}, 794 (1988)}; I. G. Knowles, \href{https://doi.org/10.1016/0550-3213(88)90092-2}{Nucl. Phys. B \textbf{310}, 571 (1988).}

 
\bibitem{Compass:rho0}
A. Kerbizi \emph{et al.} (COMPASS Collaboration), \href{https://doi.org/10.21468/SciPostPhysProc.8.146}{SciPost Phys. Proc. \textbf{8} (2022) 146.}

\end{thebibliography}
\end{document}